\documentstyle[epsfig]{aipproc}
\begin{document}
%
\pagestyle{plain}
\newcommand{\1}{{\'\i}}
\newcommand{\be}{\begin{equation}}
\newcommand{\ee}{\end{equation}\noindent}
\newcommand{\bear}{\begin{eqnarray}}
\newcommand{\ear}{\end{eqnarray}\noindent}
\newcommand{\no}{\noindent}
\date{}
\renewcommand{\theequation}{\arabic{section}.\arabic{equation}}
\renewcommand{\arraystretch}{2.5}
\newcommand{\GeV}{\mbox{GeV}}
\newcommand{\cL}{\cal L}
\newcommand{\D}{\cal D}
\newcommand{\Dhalf}{{D\over 2}}
\newcommand{\Det}{{\rm Det}}
\newcommand{\PP}{\cal P}
\newcommand{\G}{{\cal G}}
\def\R{1\!\!{\rm R}}
\def\Eins{\mathord{1\hskip -1.5pt
\vrule width .5pt height 7.75pt depth -.2pt \hskip -1.2pt
\vrule width 2.5pt height .3pt depth -.05pt \hskip 1.5pt}}
\newcommand{\symb}{\mbox{symb}}
\renewcommand{\arraystretch}{2.5}
\newcommand{\slD}{\raise.15ex\hbox{$/$}\kern-.57em\hbox{$D$}}
\newcommand{\slpartial}{\raise.15ex\hbox{$/$}\kern-.57em\hbox{$\partial$}}
\newcommand{\slG}{{{\dot G}\!\!\!\! \raise.15ex\hbox {/}}}
\newcommand{\Gd}{{\dot G}}
\newcommand{\Gund}{{\underline{\dot G}}}
\newcommand{\Gdd}{{\ddot G}}
\def\GBd12{{\dot G}_{B12}}
\def\mneg{\!\!\!\!\!\!\!\!\!\!}
\def\Mneg{\!\!\!\!\!\!\!\!\!\!\!\!\!\!\!\!\!\!\!\!}
\def\non{\nonumber}
\def\beqn*{\begin{eqnarray*}}
\def\eqn*{\end{eqnarray*}}
\def\sy{\scriptscriptstyle}
\def\footstrut{\baselineskip 12pt}
\def\square{\kern1pt\vbox{\hrule height 1.2pt\hbox{\vrule width 1.2pt
   \hskip 3pt\vbox{\vskip 6pt}\hskip 3pt\vrule width 0.6pt}
   \hrule height 0.6pt}\kern1pt}
\def\np{n_{+}}
\def\nm{n_{-}}
\def\Np{N_{+}}
\def\Nm{N_{-}}
\def\slash#1{#1\!\!\!\raise.15ex\hbox {/}}
\def\dint#1{\int\!\!\!\!\!\int\limits_{\!\!#1}}
\def\bra#1{\langle #1 |}
\def\ket#1{| #1 \rangle}
\def\vev#1{\langle #1 \rangle}
\def\rightvac{\mid 0\rangle}
\def\leftvac{\langle 0\mid}
\def\dps{\displaystyle}
\def\sy{\scriptscriptstyle}
\def\half{{1\over 2}}
\def\third{{1\over3}}
\def\fourth{{1\over4}}
\def\fifth{{1\over5}}
\def\sixth{{1\over6}}
\def\seventh{{1\over7}}
\def\eigth{{1\over8}}
\def\ninth{{1\over9}}
\def\tenth{{1\over10}}
\def\pa{\partial}
\def\ddtau{{d\over d\tau}}
\def\ge{\hbox{\textfont1=\tame $\gamma_1$}}
\def\gz{\hbox{\textfont1=\tame $\gamma_2$}}
\def\gd{\hbox{\textfont1=\tame $\gamma_3$}}
\def\go{\hbox{\textfont1=\tamt $\gamma_1$}}
\def\gt{\hbox{\textfont1=\tamt $\gamma_2$}}
\def\gth{\hbox{\textfont1=\tamt $\gamma_3$}} 
\def\gf{\hbox{$\gamma_5\;$}}
\def\ie{\hbox{$\textstyle{\int_1}$}}
\def\iz{\hbox{$\textstyle{\int_2}$}}
\def\id{\hbox{$\textstyle{\int_3}$}}
\def\ldop{\hbox{$\lbrace\mskip -4.5mu\mid$}}
\def\rdop{\hbox{$\mid\mskip -4.3mu\rbrace$}}
\def\eps{\epsilon}
\def\epshalf{{\epsilon\over 2}}
\def\e{\mbox{e}}
\def\g{\mbox{g}}
\def\kinb{{1\over 4}\dot x^2}
\def\kinf{{1\over 2}\psi\dot\psi}
\def\expk{{\rm exp}\biggl[\,\sum_{i<j=1}^4 G_{Bij}k_i\cdot k_j\biggr]}
\def\expp{{\rm exp}\biggl[\,\sum_{i<j=1}^4 G_{Bij}p_i\cdot p_j\biggr]}
\def\expshort{{\e}^{\half G_{Bij}k_i\cdot k_j}}
\def\expabb{{\e}^{(\cdot )}}
\def\epseps#1#2{\varepsilon_{#1}\cdot \varepsilon_{#2}}
\def\epsk#1#2{\varepsilon_{#1}\cdot k_{#2}}
\def\kk#1#2{k_{#1}\cdot k_{#2}}
\def\G#1#2{G_{B#1#2}}
\def\Gp#1#2{{\dot G_{B#1#2}}}
\def\GF#1#2{G_{F#1#2}}
\def\Dab{{(x_a-x_b)}}
\def\Dsq{{({(x_a-x_b)}^2)}}
\def\lag{( -\partial^2 + V)}
\def\PITD{{(4\pi T)}^{-{D\over 2}}}
\def\4piTD{{(4\pi T)}^{-{D\over 2}}}
\def\4piT4{{(4\pi T)}^{-2}}
\def\TintmD{{\dps\int_{0}^{\infty}}{dT\over T}\,e^{-m^2T}
    {(4\pi T)}^{-{D\over 2}}}
\def\Tintm4{{\dps\int_{0}^{\infty}}{dT\over T}\,e^{-m^2T}
    {(4\pi T)}^{-2}}
\def\Tintm{{\dps\int_{0}^{\infty}}{dT\over T}\,e^{-m^2T}}
\def\Tint{{\dps\int_{0}^{\infty}}{dT\over T}}
\def\pint{{\dps\int}{dp_i\over {(2\pi)}^d}}
\def\Dx{\dps\int{\cal D}x}
\def\Dy{\dps\int{\cal D}y}
\def\Dpsi{\dps\int{\cal D}\psi}
\def\Tr{{\rm Tr}\,}
\def\tr{{\rm tr}\,}
\def\sumij{\sum_{i<j}}
\def\freeexp{{\rm e}^{-\int_0^Td\tau {1\over 4}\dot x^2}}
\def\arraystretch{2.5}
\def\Ge{\mbox{GeV}}
\def\dA{\partial^2}
\def\DA{\sqsubset\!\!\!\!\sqsupset}
\def\FFdual{F\cdot\tilde F}
\def\bbbr{{\rm I\!R}}
\def\bbbone{{\mathchoice {\rm 1\mskip-4mu l} {\rm 1\mskip-4mu l}
{\rm 1\mskip-4.5mu l} {\rm 1\mskip-5mu l}}}
\def\bbbz{{\mathchoice {\hbox{$\sf\textstyle Z\kern-0.4em Z$}}
{\hbox{$\sf\textstyle Z\kern-0.4em Z$}}
{\hbox{$\sf\scriptstyle Z\kern-0.3em Z$}}
{\hbox{$\sf\scriptscriptstyle Z\kern-0.2em Z$}}}}
%
\renewcommand{\theequation}{\arabic{equation}}
\setcounter{equation}{0}
\title{QED in the Worldline Formalism
\footnote{Talk given at QED 2000, 2nd workshop on
``Frontier Tests of Quantum Electrodynamics 
and Physics of the Vacuum'', Trieste, Italy,
Oct. 5-11, 2000.}
}
\author{Christian Schubert}
\address{Instituto de F\1sica y Matem\'aticas,
Universidad Michoacana,
Morelia, Michoac\'an, M\'exico
\\
and\\
California Institute for Physics and Astrophysics,
366 Cambridge Ave., Palo Alto, California}
\maketitle
\begin{abstract}
A survey is given of applications of the ``string-inspired''
worldline formalism to the computation of amplitudes and effective
actions in QED. 
\end{abstract}

\section*{Introduction: QED in First Quantization}

The ``worldline'' or ``string-inspired'' formalism is an
alternative to the usual second-quantized formalism in
quantum field theory based on relativistic particle path integrals.
Although it was invented by Feynman \cite{feynman1,feynman2}
simultaneously with modern relativistic second-quantized  
QED, until recently it was only
occasionally used for actual computations \cite{early}. 
During the last few
years, however, certain computational advantages of the first-quantized
approach were recognized which led to a sizeable number of
nontrivial applications. 

Those recent developments were triggered by the work of Bern and
Kosower, who derived new rules for the construction of one-loop
QCD amplitudes from the infinite string tension limit of
first-quantized string path integrals \cite{berkos}.
Strassler 
then showed \cite{strassler}
that the corresponding formulas for
the QED case can also be derived from 
Feynman's first quantized path integrals.

For the case of scalar QED, the basic formula is given in the
appendix A of \cite{feynman1}. 
It states that
the amplitude for a charged scalar particle to move,
under the influence of the external potential $A_{\mu}$,
from point $x_{\mu}$ to $x_{\mu}'$ in Minkowski space
is given by

\begin{figure}[ht]
\vbox to 4.5cm{\vfill\hbox to 15.8cm{\hfill
\epsffile{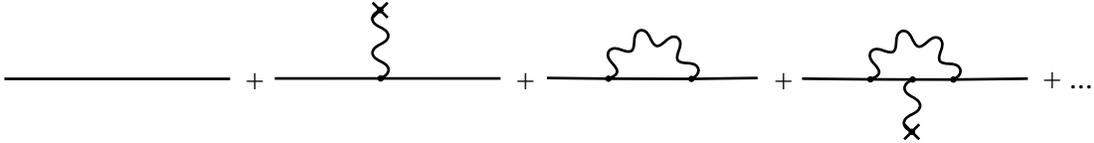}
\hfill}\vfill}\vskip-.4cm
\caption[dum]{Sum of Feynman diagrams represented by a single
path integral.
\hphantom{xxxxxxxxxxxxxxx}}
\label{totalelectronprop}
\end{figure}

\bear
\int_0^{\infty}
ds
\int
_{x(0)=x}^{x(s)=x'}
{\cal D}x(\tau)
&&
\exp
\Bigl(
-{1\over 2}
im^2s\Bigr)
\exp
\biggl[
-{i\over 2}
\int_0^sd\tau
{({dx_{\mu}\over d\tau})}^2
-i\int_0^s
d\tau
{dx_{\mu}\over d\tau}
A_{\mu}(x(\tau))\nonumber\\
&&
-{i\over 2}
e^2
\int_0^sd\tau
\int_0^sd\tau'
{dx_{\mu}\over d\tau}
{dx_{\nu}\over d\tau'}
\delta_{+}^{\mu\nu}(x(\tau)-x(\tau'))
\biggr]
\label{feynform}
\ear
That is,
the amplitude can be constructed as a 
path integral over the set of all open
trajectories running from $x$ to $x'$ in the fixed
proper time $s$. The action 
consists of the familiar kinetic term, and two interaction terms.
Of those the first represents the interaction with the external field,
to all orders in the field, while the second one describes an
arbitrary number of virtual photons emitted and re-absorbed along the
trajectory of the particle.  Here $\delta_{+}^{\mu\nu}$ denotes the photon
propagator, e.g. in Feynman gauge,
$\delta_{+}^{\mu\nu}(x,x') = {g^{\mu\nu}\over (x-x')^2}$. 
In second quantized field theory this amplitude would thus
correspond to the infinite sequence of Feynman diagrams
shown in fig. \ref{totalelectronprop}.

\par
\par

\noindent

Feynman showed that, starting from this basic formula, the complete 
scalar QED S-matrix
can be constructed by writing a separate path integral for
every scalar line, open or closed, and interconnecting
them by further photon insertions in all possible
ways. One year later he extended this work to the
case of spinor QED \cite{feynman2}; however, at the time
no satisfactory way of computing these path integrals was found,
and the formalism seems to have had no impact on the 
further development of quantum field theory.

\section*{$N$ -- Photon Amplitudes}

We will now explain the ``string-inspired'' way of computing
this type of path integral. Only photon amplitudes will
be considered in the present talk; while the method applies
also to 
external scalars/electrons \cite{extscalel},
too few nontrivial computations of such amplitudes have been
done yet to judge the usefulness of this extension. 
Let us start with the simplest case, one-loop $N$-photon
scattering in scalar QED.
Rewritten for the case of a closed loop, and not taking
internal photon corrections into account, Feynman's
formula (\ref{feynform}) turns into a representation of
the one-loop effective action for the Maxwell
field:

\begin{equation}
\Gamma\lbrack A\rbrack = \int_0^{\infty}
{dT\over T} \, {\rm e}^{-m^2T}
\int {\cal D}x\, 
{\rm exp} \left[ 
- \int_0^T \!\!\! d\tau \left( {1\over 4}{\dot x}^2 
+ ieA_{\mu}\dot x^{\mu} 
\right) \right]
\label{scalarpi}
\end{equation}
\no
The path integral runs now over the space of closed trajectories
with period $T$, $x^{\mu}(T)=x^{\mu}(0)$ 
\footnote{
The proper time
parameter
$s$ has been rescaled and Wick rotated,
$s\rightarrow -i2T$.}.
If we expand the
``interaction exponential'',

\be
{\rm exp}\Bigl[
-\int_0^Td\tau\, ieA_{\mu}\dot x^{\mu}
\Bigr]
=\sum_{N=0}^{\infty}
{{(-ie)}^N\over N!}
\prod_{i=0}^N
\int_0^Td\tau_i
\biggl[
\dot x^{\mu}(\tau_i)
A_{\mu}(x(\tau_i))
\biggr]
\label{expandint}
\ee\no
the individual terms correspond to Feynman diagrams
describing a fixed number of
interactions of the scalar loop with
the external field.
The corresponding $N$ -- photon
scattering amplitude is then obtained by
specializing to a background
consisting of 
a sum of plane waves with definite
polarizations,
$
A_{\mu}(x)=
\sum_{i=1}^N
\varepsilon_{i\mu}
\e^{ik_i\cdot x}
$,
and picking out the term containing every
$\varepsilon_i $ once.
This yields the following representation of
the $N$ - photon amplitude, 

\bear
\Gamma[\lbrace k_i,\varepsilon_i\rbrace\rbrace]
&=&
(-ie)^{N}
\Tintm 
\int {\cal D}x\,
V_{\rm scal}^{A}[k_1,\varepsilon_1]\ldots
V_{\rm scal}^{A}[k_N,\varepsilon_N]
\e^{-\int_0^Td\tau {\dot x^2\over 4}}
\non\\
\label{repNvector}
\ear\no
Here $V_{\rm scal}^{A}$ denotes 
the same photon
vertex operator as is used in string perturbation
theory,
$
V_{\rm scal}^{A}[k,\varepsilon]
\equiv
\int_0^Td\tau\,
\varepsilon\cdot \dot x(\tau)
\,{\rm e}^{ikx(\tau)}
$.
At this stage the path integral has become Gaussian,
which in principle reduces its evaluation to the 
determination of the appropriate Green's function.
However a zero-mode given by the constant
loops must be removed first, which
is done by fixing the average position
$x_0^{\mu}= {1\over T}\int_0^T d\tau\, x^{\mu}(\tau)$
of the loop. 
In scattering amplitude calculations
the integral over $x_0$ factors out
and produces the usual momentum conservation factor
$(2\pi)^D\delta(\sum k_i)$. 
The reduced path integral $\int{\cal D}y(\tau)$
over $y(\tau)\equiv x(\tau) - x_0$
can then be evaluated using the ``bosonic''
worldline Green's function $G_B$,

\bear
\langle y^{\mu}(\tau_1)y^{\nu}(\tau_2)\rangle
= -g^{\mu\nu}G_B(\tau_1,\tau_2)=
-g^{\mu\nu}\Bigl[
\mid \tau_1-\tau_2\mid 
-{{(\tau_1-\tau_2)}^2\over T}
\Bigr] 
\label{defGB}
\ear
\no
Using a formal exponentiation of the factors
$\varepsilon_i\cdot\dot x_i$'s, and ``completing the
square'', one arrives at the following
closed expression for the one-loop
$N$ - photon amplitude:

\begin{eqnarray}
\Gamma[\lbrace k_i,\varepsilon_i\rbrace]
&=&
{(-ie)}^N
{(2\pi )}^D\delta (\sum k_i)
{\dps\int_{0}^{\infty}}{dT\over T}
{(4\pi T)}^{-{D\over 2}}
e^{-m^2T}
\prod_{i=1}^N \int_0^T 
d\tau_i
\nonumber\\
&&\hspace{-80pt}
\times
\exp\biggl\lbrace\sum_{i,j=1}^N 
\bigl\lbrack \half G_{Bij} k_i\cdot k_j
+i\dot G_{Bij}k_i\cdot\varepsilon_j 
+\half\ddot G_{Bij}\varepsilon_i\cdot\varepsilon_j
\bigr\rbrack\biggr\rbrace
\mid_{\rm multi-linear}
\nonumber\\
\label{scalarqedmaster}
\end{eqnarray}
\no
Here it is understood that only the terms linear
in all the $\varepsilon_1,\ldots,\varepsilon_N$
have to be taken. 
Besides the Green's function $G_B$ also its first and
second deriatives appear,
$\dot G_B(\tau_1,\tau_2) = {\rm sign}(\tau_1 - \tau_2)
- 2 {{(\tau_1 - \tau_2)}\over T},
\ddot G_B(\tau_1,\tau_2)
= 2 {\delta}(\tau_1 - \tau_2)
- {2\over T}$.
Dots generally denote a
derivative acting on the first variable,
and we abbreviate
$G_{Bij}\equiv G_B(\tau_i,\tau_j)$ etc.
The factor ${(4\pi T)}^{-{D\over 2}}$
represents the free Gaussian path integral
determinant.
The expression (\ref{scalarqedmaster})
is identical with the ``Bern-Kosower
Master Formula'' 
for the special case considered \cite{berkos}. 

In the spinor QED case, the path integral used in the worldline
formalism is somewhat different from Feynman's original proposal
\cite{feynman2}, which involved Dirac matrices. The modern
version encodes the Clifford algebra combinatorics 
in an additional path integral ${\cal D}\psi$,
representing the electron spin, which runs over the
space of Grassmann functions $\psi^{\mu}(\tau)$
antiperiodic in the proper-time
\cite{fradkin}.
This allows one to write
the spinor loop equivalent of (\ref{scalarpi}) in a formally
analogous way \cite{superloop}:

\begin{equation}
\Gamma_{\rm spin}
\lbrack A\rbrack   = - \half{\displaystyle\int_0^{\infty}}
{dT\over T}
e^{-m^2T}
{\displaystyle\int} {\cal D} X
{\rm e}^{-
\int_0^T d\tau\int d\theta \, \Bigl [-{1\over 4}X\cdot D^3 X 
- ieDX\cdot A(X)\Bigr ]}
\label{superpi}
\end{equation}
\no
where
$X^{\mu} = x^{\mu} 
+ \sqrt 2\,\theta\psi^{\mu}, 
Y^{\mu} = X^{\mu}-x_0^{\mu},
D = {\partial\over{\partial\theta}} - 
   \theta
{\partial\over{\partial\tau}},
\int d\theta\theta = 1$. 

Up to normalization
this super path integral agrees
with the scalar path integral above if
all Grassmann terms are dropped.
The appropriate correlator for the evaluation of the
additional Grassmann path integral
is $\langle \psi(\tau_1)\psi(\tau_2)\rangle = 
\half g^{\mu\nu}G_F(\tau_1,\tau_2)$, with 
$G_F(\tau_1,\tau_2) = {\rm sign}(\tau_1-\tau_2)$.
Its explicit evaluation can, however, be circumvented,
using the following ``replacement rule'' found by
Bern and Kosower:

Writing out the exponential in the master formula
eq.(\ref{scalarqedmaster}) for a fixed number $N$ of
photons, one obtains an integrand

\be \exp\biggl\lbrace 
\biggr\rbrace \mid_{\rm multi-linear} \quad={(-i)}^N P_N(\dot
G_{Bij},\ddot G_{Bij}) \exp\biggl[\half \sum_{i,j=1}^N G_{Bij}k_i\cdot
k_j \biggr] \label{defPN} \ee\no 
with a certain polynomial $P_N$
depending on the various  $\dot G_{Bij}$'s, $\ddot G_{Bij}$'s, 
as well as on the kinematic invariants. $P_N$ corresponds
to the Feynman numerator in a Feynman parameter
integral calculation; the term in the exponent 
turns, after the execution of the
global proper-time integral, into the standard Feynman denominator.
By suitable partial integrations all second derivatives
$\ddot G_{Bij}$ appearing in $P_N$ can be removed,
so that $P_N$ gets replaced by
another polynomial $Q_N$ depending solely
on the $\dot G_{Bij}$'s,

\be
P_N(\dot G_{Bij},\ddot G_{Bij})
\,\e^{\half\sum G_{Bij}k_i\cdot k_j}
\quad
{\stackrel{\sy{\rm part. int.}}{\longrightarrow}}
\quad
Q_N(\dot G_{Bij})
\,\e^{\half\sum G_{Bij}k_i\cdot k_j}
\label{partint}
\ee\no
Then the complete integrand for the spinor loop
case can be obtained by simultaneously
replacing every closed cycle 
$\dot G_{Bi_1i_2}\dot G_{Bi_2i_3}\cdots\dot G_{Bi_ki_1}$
appearing in $Q_N$ by

\begin{equation}
\dot G_{Bi_1i_2}
\dot G_{Bi_2i_3}\cdots\dot G_{Bi_ki_1}
- G_{Fi_1i_2}G_{Fi_2i_3}\cdots
G_{Fi_ki_1}
\label{fermion}
\end{equation}\no
(and multiplying by a global factor of $-2$). 
The result of the partial integration 
procedure is not unique, but can be made so
by requiring $Q_N$ to have the full permutation
symmetry in the $N$ photons \cite{menphoton}.
As a bonus of the partial integration procedure, one
obtains a maximal gauge invariant decomposition of
the $N$ - photon amplitudes \cite{menphoton}.
Up to the six-point case, the polynomials $Q_N$
can still be written quite compactly, and
are given explicitly in \cite{menphoton}.
The usefulness of $Q_6$ for the computation of the
six-photon amplitudes is presently under 
investigation.

\section*{$N$ -- Photon Amplitudes in Constant Fields}

In this formalism, the inclusion of constant external fields 
requires only relatively minor modifications
\cite{ss1,cadhdu,shaisultanov,rescsc}.
For this reason it has been extensively applied
to constant field processes in QED 
\cite{ss1,cadhdu,rescsc,gussho,adlsch,frss,korsch,ditsha,mevv}.

Let us thus introduce an additional
background field $\bar A^{\mu}(x)$
with constant field strength tensor
$\bar F_{\mu\nu}$. 
Using Fock--Schwinger gauge
centered at the loop average position $x_0$ \cite{ss1} we may
take $\bar A^{\mu}(x)$ to be of the form
$\bar A_{\mu}(x) = 
{1\over 2}y^{\nu}\bar F_{\nu\mu}$.
The constant field contribution to the 
worldline Lagrangian in the spinor QED path integral (\ref{superpi})
then becomes
$\Delta L = -{1\over 2}ieY^{\mu}\bar F_{\mu\nu}DY^{\nu}$.
Since it is quadratic in the worldline
fields, it can be taken into account by appropriate
changes of the worldline propagators and the path integral
determinant. Those are (deleting the ``bar'')

\begin{eqnarray}
G_{B12}&\rightarrow&{\cal G}_{B12}\equiv
{T\over 2{({\cal Z})}^2}\biggl({{\cal Z}\over{{\rm sin}({\cal Z})}}
{\rm e}^{-i{\cal Z}\dot G_{B12}}
+i{\cal Z}\dot G_{B12} -1\biggr)
\nonumber\\
G_{F12}&\rightarrow&{\cal G}_{F12} =
G_{F12}
{{\rm e}^{-i{\cal Z}\dot G_{B12}}\over {\rm cos}({\cal Z})}
\nonumber\\
{(4\pi T)} ^{-{D\over 2}}
&\rightarrow&
{(4\pi T)}^{-{D\over 2}}
{\rm det}^{-{1\over 2}}
\biggl[{\sin({\cal Z})\over {{\cal Z}}}
\biggr] \qquad {(\rm Scalar\quad QED)}
\nonumber\\
{(4\pi T)} ^{-{D\over 2}}
&\rightarrow&
{(4\pi T)}^{-{D\over 2}}
{\rm det}^{-{1\over 2}}
\biggl[{\tan({\cal Z})\over {{\cal Z}}}
\biggr] \qquad {(\rm Spinor\quad QED)}
\nonumber\\
\label{calGBGFdet}
\end{eqnarray}
\noindent
These expressions should be understood as power
series in the Lorentz matrix ${\cal Z}^{\mu\nu}\equiv eTF^{\mu\nu}$.

Retracing the above evaluation of the $N$ - photon amplitude in
vacuum one obtains the following generalization of 
(\ref{scalarqedmaster}),
representing the scalar QED $N$ - photon scattering amplitude 
in a constant field \cite{shaisultanov,rescsc}:
\vspace{5pt}
\begin{eqnarray}
&&\Gamma_{\rm scal}
[\lbrace k_i,\varepsilon_i\rbrace]
=
{(-ie)}^N
{(2\pi )}^D\delta (\sum k_i)
{\dps\int_{0}^{\infty}}{dT\over T}
{(4\pi T)}^{-{D\over 2}}
e^{-m^2T}
{\rm det}^{-{1\over 2}}
\biggl[{{\rm sin}({\cal Z})\over {\cal Z}}\biggr]
\label{scalarqedmasterF}\\
&&\hspace{20pt}\times
\prod_{i=1}^N \int_0^T 
d\tau_i
\exp\biggl\lbrace\sum_{i,j=1}^N 
\Bigl\lbrack \half k_i\cdot {\cal G}_{Bij}\cdot  k_j
-i\varepsilon_i\cdot\dot{\cal G}_{Bij}\cdot k_j
+\half
\varepsilon_i\cdot\ddot {\cal G}_{Bij}\cdot\varepsilon_j
\Bigr\rbrack\biggr\rbrace
\mid_{\rm multi-linear}\quad
\nonumber
\end{eqnarray}
\vspace{5pt}

\no
The transition from the scalar to the spinor QED
case proceeds in complete analogy to the above, 
except for the fact that in the external field case the
path integral determinants are different in both cases.
The cycle replacement rule (\ref{fermion}) remains valid
{\sl mutatis mutandis}.

For example, in the vacuum polarization case $N=2$
eq.(\ref{scalarqedmasterF})
yields, after partial integration 
and application of the
``cycle replacement rule'', the following
integrand for the spinor QED vacuum polarization
tensor in a constant field,

\bear
I^{\mu\nu} &=& \e^{-k\cdot ({\cal G}_{B12}-{\cal G}_{B11})\cdot k}
\biggl\lbrace
\dot{\cal G}^{\mu\nu}_{B12}k\cdot\dot{\cal G}_{B12}\cdot k
-{\cal G}^{\mu\nu}_{F12}k\cdot{\cal G}_{F12}\cdot k
\non\\&&\hspace{-15pt}
- \biggl[
\Bigl(\dot{\cal G}_{B11}
-{\cal G}_{F11}
-\dot{\cal G}_{B12}\Bigr)
^{\mu\lambda}
  \Bigl(\dot{\cal G}_{B21}
-\dot{\cal G}_{B22}
+{\cal G}_{F22}
\Bigr)^{\nu\kappa}
+
{\cal G}^{\mu\lambda}_{F12}
{\cal G}^{\nu\kappa}_{F21}
\biggr]
k^{\kappa}k^{\lambda}
\biggr\rbrace
\non\\
\label{substint}
\ear\no
Specializing to a Lorentz system 
where ${\bf E}=(0,0,E)$ and ${\bf B}=(0,0,B)$,
the final (Minkowski space) result for the on-shell
renormalized scalar and spinor
QED vacuum polarization tensors in a constant field
looks as follows \cite{mevv},

\bear
{\bar \Pi}^{\mu\nu}_{\bigl({{\rm spin}\atop{\rm scal}}\bigr)}(k)
&=&
-{\alpha\over 4\pi}
\biggl({-2\atop 1}\biggr)
{\dps\int_{0}^{\infty}}{ds\over s}
\int_{-1}^1 {dv\over 2}
\Biggl\lbrace
{zz'\over \sin(z)\sinh(z')}
\biggl({\cos(z)\cosh(z')\atop 1}\biggr)
\,\e^{-is\Phi_0}
\non\\
&&\hspace{-50pt}
\times
\sum_{\alpha,\beta =\perp ,\parallel}
\Bigl[
s^{\alpha\beta}_{\bigl({{\rm spin}\atop{\rm scal}}\bigr)}
({\eta}^{\mu\nu}_{\alpha}k_{\beta}^2-k^{\mu}_{\alpha}k^{\nu}_{\beta})
+a^{\alpha\beta}_{\bigl({{\rm spin}\atop{\rm scal}}\bigr)}
\tilde k_{\alpha}^{\mu}\tilde k_{\beta}^{\nu}\Bigr]
-\e^{-ism^2}({\eta}^{\mu\nu}k^2-k^{\mu}k^{\nu})
\biggl({v^2-1\atop v^2}\biggr)
\Biggr\rbrace
\label{specialvpfinal}
\ear\no
Here $k_{\parallel} = (k^0,0,0,k^3),
k_{\perp} = (0,k^1,k^2,0),
\tilde k_{\parallel} = (k^3,0,0,k^0),
\tilde k_{\perp} = (0,k^2,-k^1,0)$,
and the coefficient functions 
$\Phi_0,s_{\rm scal/spin}^{\alpha\beta},a_{\rm scal/spin}^{\alpha\beta}$
involve trigonometric functions of the two
variables $z=eBs,\, z'=eEs$.
This result can be shown to agree
with the field theory results of 
\cite{scalvp} (scalar QED) and 
\cite{spinvp} (fermion QED). 
A somewhat different version of the same worldline calculation was
presented in \cite{ditsha}. 
See \cite{adlsch} for an application of (\ref{scalarqedmasterF})
to the recalculation of the scalar/spinor QED 
amplitudes for photon splitting in a constant magnetic field.

\section*{Euler-Heisenberg Type Lagrangians}

Let us look more closely at the photonless case, i.e. the vacuum 
amplitude in a constant field. At the one-loop level
this quantity is described by the
well-known Euler-Heisenberg-Schwinger Lagrangians.
In the present formalism,
these Lagrangians 
are essentially given by the determinant factors
introduced above. For example, the scalar QED
path integral (\ref{scalarpi}) turns, for a constant
field in Fock-Schwinger gauge, into an easily computable
Gaussian path integral,

\bear
\Gamma\lbrack F\rbrack &=& \int_0^{\infty}
{dT\over T} \, {\rm e}^{-m^2T}
\int {\cal D}x\, 
{\rm exp} \left[ 
- \int_0^T \!\!\! d\tau \left( {1\over 4}{\dot x}^2 
+ {i\over 2}ex^{\mu}F_{\mu\nu}\dot x^{\nu} 
\right) \right]
\non\\
&=& \int_0^{\infty}
{dT\over T} \, {\rm e}^{-m^2T}
\int dx_0
\,
{\rm Det'}^{-{1\over 2}}_P
\Bigl[ -{d^2\over d\tau^2}\Bigr]
{\rm Det'}^{-{1\over 2}}_P
\Bigl[\Eins 
-2ieF\Bigl({d\over d\tau}\Bigr)^{-1}\Bigr]
\non\\
&=& \int_0^{\infty}
{dT\over T} \, {\rm e}^{-m^2T}
\int dx_0
\,(4\pi T)^{-{D\over 2}}
{\rm det}^{-{1\over 2}}
\biggl[{\sin(eFT)\over eFT}\biggr] 
\non\\
&=& -{1\over 16\pi^2}\int dx_0\int_0^{\infty}{ds\over s}
\,\e^{-ism^2}\,
{e^2ab\over \sin(eas)\sinh(ebs)}
\label{eulheiscal}
\ear\no
where $a^2-b^2 \equiv {\bf B}^2 - {\bf E}^2$,
$ab \equiv {\bf E}\cdot{\bf B}$. The last line
is Schwinger's result \cite{schwinger51}.

The two-loop generalization of the Euler-Heisenberg-Schwinger
formulas, involving a single photon insertion into the
scalar/electron loop, 
was obtained decades ago by Ritus \cite{ritus}. 
In \cite{rescsc,frss} the case of the two-loop effective
action in a purely magnetic constant field was taken up again
in the worldline formalism, both in scalar and spinor QED.
Following \cite{ss3} in those
calculations the denominator of the insertion term
 
$$\int_0^T d\tau_a \int_0^T d\tau_b
{\dot x(\tau_a)\cdot\dot x(\tau_b)\over
(x(\tau_a) - x(\tau_b))^2}$$

\noindent
in (\ref{feynform}), or its super analogue,
was exponentiated through the introduction
of a photon proper-time integral. The total
path integral is then still Gaussian, and can be computed
by Wick contractions using effective worldline
propagators taking both the external field
and the photon insertion into account.
The two-loop (on-shell renormalized) 
effective action was then obtained in form
of a two-fold integral over $T$ and the relative position
of the photon endpoints on the loop. Those integrals cannot
be done in closed form, but the coefficients of the
effective Lagrangian as a weak field expansion in $B^2$,

\begin{eqnarray}
{\cal L}^{(2)} = \alpha\, \frac{m^4}{(4\pi)^3} \left(\frac{eB}{m^2}\right)^4  
\sum_{n=0}^\infty a_n^{(2)} \, \left(\frac{eB}{m^2}\right)^{2n}
\label{2leh}
\end{eqnarray}
can be computed to any desired order, and it was in this
way that the equivalence to the results of \cite{ritus} was
established. The generalization of the worldline
calculation to a general constant field was done in
\cite{korsch}.

More recently we have, for fermion QED, extended this work 
to the case of a constant electrical field \cite{dunsch}. 
Although this case can be obtained from
the magnetic one by a trivial substitution of
$B^2\rightarrow -E^2$, 
it is qualitatively different insofar as
only in the presence of the electric field the
effective action acquires a
nonperturbative imaginary part, indicative
of the possibility of pair creation \cite{schwinger51}.
The imaginary part of the two-loop effective action was
studied by Lebedev and Ritus \cite{lebrit}. Adding their two-loop
result to Schwinger's corresponding one-loop result 
\cite{schwinger51} one has

\begin{eqnarray}
{\rm Im} {\cal L}^{(1)} +
{\rm Im} {\cal L}^{(2)} &\sim&  \frac{m^4}{8\pi^3}
\beta^2\,
\sum_{k=1}^\infty
\Bigl[
\frac{1}{k^2}
+\alpha\pi K_k(\beta)
\Bigr]
\,\exp\left[-\frac{\pi k}{\beta}\right]
\label{fullimag2loop}
\end{eqnarray}
where $\beta = {eE\over m^2}$. The function $K_k(\beta)$
has the following small $\beta$ - expansion, 

\begin{eqnarray}
K_k(\beta) &=& -{c_k\over \sqrt{\beta}} + 1 + {\rm O}(\sqrt{\beta}),
\quad
c_1 = 0,\quad
c_k = {1\over 2\sqrt{k}}
\sum_{l=1}^{k-1} {1\over \sqrt{l(k-l)}},
\quad k \geq 2
\nonumber\\
\label{expK}
\end{eqnarray}
In \cite{lebrit} this was shown by an analysis of the analyticity
properties of the parameter integrals. 
In \cite{dunsch} a very different approach was taken to the construction
of the imaginary part, based on a Borel summation of the weak field
expansion. The method itself will only be sketched here, since it is discussed
in detail in G. Dunne's talk at this conference \cite{dunnetalk}.

Using MATHEMATICA we obtained, from the
parameter integral representation given in
\cite{frss}, the expansion coefficients $a_n^{(2)}$ in
(\ref{2leh}) up to $n=15$. This turned out to
be sufficient for a reliable fit of the leading and subleading
asymptotic behaviour of the series, of the following form:

\begin{eqnarray}
a_n^{(2)}\sim (-1)^n \frac{16}{\pi^2} \, \frac{\Gamma(2n+2)}{\pi^{2n}}\left[  
1-\frac{0.44}{\sqrt{n}}+\dots \right]
\label{2lcorr}
\end{eqnarray}
Due to the alternating behaviour of the series this implies that the
weak field expansion is Borel summable in the magnetic case.
The expansion coefficients for the electric field case differ from the
magnetic ones by the absence of the factor $(-1)^n$, which 
destroys the Borel summability.
However, the Borel sum of the alternating magnetic series can be used to
construct, by a dispersion relation, the
imaginary part of the corresponding non-alternating sum. In this
way we  rederived from (\ref{2lcorr}) the leading term 
in the expansion of the function $K_1(\beta )$ appearing
in (\ref{fullimag2loop}), and obtained a numerical value for
its first subleading correction.
Adding those two terms 
to the leading term in Schwinger's one-loop formula
gives the following,

\begin{eqnarray}
{\rm Im} \left({\cal L}^{(1)}+{\cal L}^{(2)}\right) \sim
\left(1+\alpha\,\pi\left[1- (0.44) \sqrt{\frac{2eE}{\pi
m^2}}+\dots\right]\right)\, \frac{m^4}{8\pi^3}\left(\frac{eE}{m^2}\right)^2\,   
\exp\left[-\frac{m^2\pi}{eE}\right]
\label{nextsum}
\end{eqnarray}
This means that, at the two-loop level, the subleading correction is much less
suppressed than at the one-loop level. It would be of obvious interest to
know the asymptotic behaviour of the higher order corrections, however to
get reliable information on them one would have to compute a much larger 
number of coefficients. It is interesting to observe, though, that
the pure magnetic or electric constant fields are not the simplest ones
which one can study in this context. The simplest, albeit unphysical, case
is the one of a self-dual Euclidean field, defined by
$F_{\mu\nu}=\half\varepsilon_{\mu\nu\alpha\beta}F^{\alpha\beta}$. 
The self-duality implies that
$F^2 = -f\Eins$
where $f=\fourth F_{\mu\nu}F^{\mu\nu}$.
Here the integrands of the
two-loop Euler-Heisenberg formulas become so simple that the second
parameter integral can be performed exactly, leaving only the
global proper-time integral. Let us write down the final result for
the on-shell renormalized
self-dual two-loop effective Lagrangian in (Euclidean)
scalar QED 
\footnote{Here we omit the
irrelevant subtraction terms from wave function renormalization.}
\cite{wop}:

\bear
{\cal L}^{(2)}_{\rm scal}[f]&=& 
{e^2\over(4\pi)^4}
\int_0^{\infty}{dT\over T^3}\,\e^{-m^2T}
\Bigl({Z\over\sinh(Z)}\Bigr)^2
\biggl\lbrace
-4Z^2
+3m^2T\Bigl[1-\gamma +\ln\Bigl({Z\over\sinh(Z)m^2T}\Bigr)
\Bigr]
\biggr\rbrace
\non\\
\label{Lfinal}
\ear
($Z\equiv \sqrt{f}eT$). 
Thus for this case it is considerably easier to obtain information on the
expansion coefficients, making it possible to
push this analysis further \cite{wop}. 

\section*{Generalizations}

Finally, let me at least mention a few generalizations
of relevance for QED which could not be discussed here:
The worldline formalism has been extended to the finite temperature
case in QED, both to the computation of the $N$ - photon amplitude
\cite{sato} and of the QED effective action \cite{shovkovy}.
Other generalizations allow one to compute
vector-axialvector \cite{newax} and vector-pseudoscalar \cite{oldax}
amplitudes, and thus photon-neutrino or photon-axion processes.

\end{document}